# PHOTOMETRY DIFFERENTIAL OF COMETS

T. Scarmato[1]

*1 - Toni Scarmato's Observatory, San Costantino di Briatico, Calabria, Italy*

*Version 1 – 2014 September 6*

## ABSTRACT

Here we present the results obtained with three comets **73P/Schwassmann-Wachmann B** the component B, **6P/d'Arrest** and **C/2008 J1 (Boattini)**. Comet 73P was observed in April 2006 with a CCD camera Starlight Express SXL8 attached to direct heat of a Newton telescope 250 mm, focal length 1200 mm, while the 6P and C/2008 J1 were ever observed by telescope Newtonian but with a CCD camera Atik 16IC to direct heat. To the CCD camera was added a filter photometric Rc (Schuler). For 6P comet has been used a defocus of 14 pixels, meaning that the central false nucleus has defocused a FWHM of 14 pixels. The technical data of the tools used are shown in Tables 1 and 2.

Keys words: General; comets, photometry.

## INTRODUCTION

Over the past 10 years the survey conducted on Earth by comets is that space has revealed that these vagabonds of the area have unique characteristics and not all behave the same way and not all have the same composition. It 'been discovered that in addition to the "classical" code of ions and dust, comets may produce code of sodium and lately has been observed in comet C/2006 P1 an **IRON NEUTRAL TAIL**. (**Fulle et al. 2007**; See the link http://digilander.libero.it/infosis/homepage/astronomia/c2006p1.htm)

One of the ways, and the scope of instruments amateur, to observe comets is the visual observations. Depending on the brightness of the comet (magnitude) just a simple binoculars to be able to admire the comets. Until a few years ago the dedicated amateur astronomers are so profitable enough to estimate the magnitude of visual comets, which can be done based on certain methods. One of these, described simply, is to defocus the coma of the comet to make the surface uniform brightness and then if in the instrument field there are some stars brighter and fainter since the comet, comparing the brightness of the comet with that of the stars defocused that you know the magnitude and then make an estimate of the apparent magnitude of the comet. (See ICQ - http://www.cfa.harvard.edu/icq/icq.html for more information). These estimates and allow themselves to still provide the photometric evolution of the comet, even though the reserve of comets with beautiful surprises "outburst" (increases brightness) that evolve both short and medium and long term. This suggests that comets are bodies active AND RETAIN and probably even when they seem really quiet in the area continues to issue material in the coma, which then expands into space, ALSO AT GREAT DISTANCE FROM SUN.

One of the most interesting is that of a comet **29P** a periodic comet showing a very interesting activities for more than 6 U. A. from the Sun with outburst that could be periodic. Again, however, the mechanism underlying its behaviour has not been explained satisfactorily and scientifically certified.

The issuance of this material change continuously over time in terms of emission od dust and gas and thus its brightness. We can ask; It can observe the changes in brightness even when they are small? And what is the accuracy of the measures that need to be able to have a session in observation of hours detect variations in brightness?

It is possible with amateur instruments to reach levels of accuracy in measuring the magnitude and possibly other parameters to monitor in the short to medium term evolution of photometric parameters of a comet or, if this is far from the sun and "apparently" inactive, the brightness variation of the "nucleus"?

To answer these questions may come to the aid **PHOTOMETRY DIFFERENTIAL**. The differential photometry, applied to the transit of extrasolar planets, has shown that with amateur instruments can reach accuracies thousandths of magnitude. There are many transits observed with this technique and a very difficult transit was discovered by amateur instruments. This is HD 17156b (Barbieri et al. 2007), whose brightness variation is only magnitudes 0.005. (In Fig. 1 you can see the change in the brightness of a star that is suspect the existence of a planet transit).

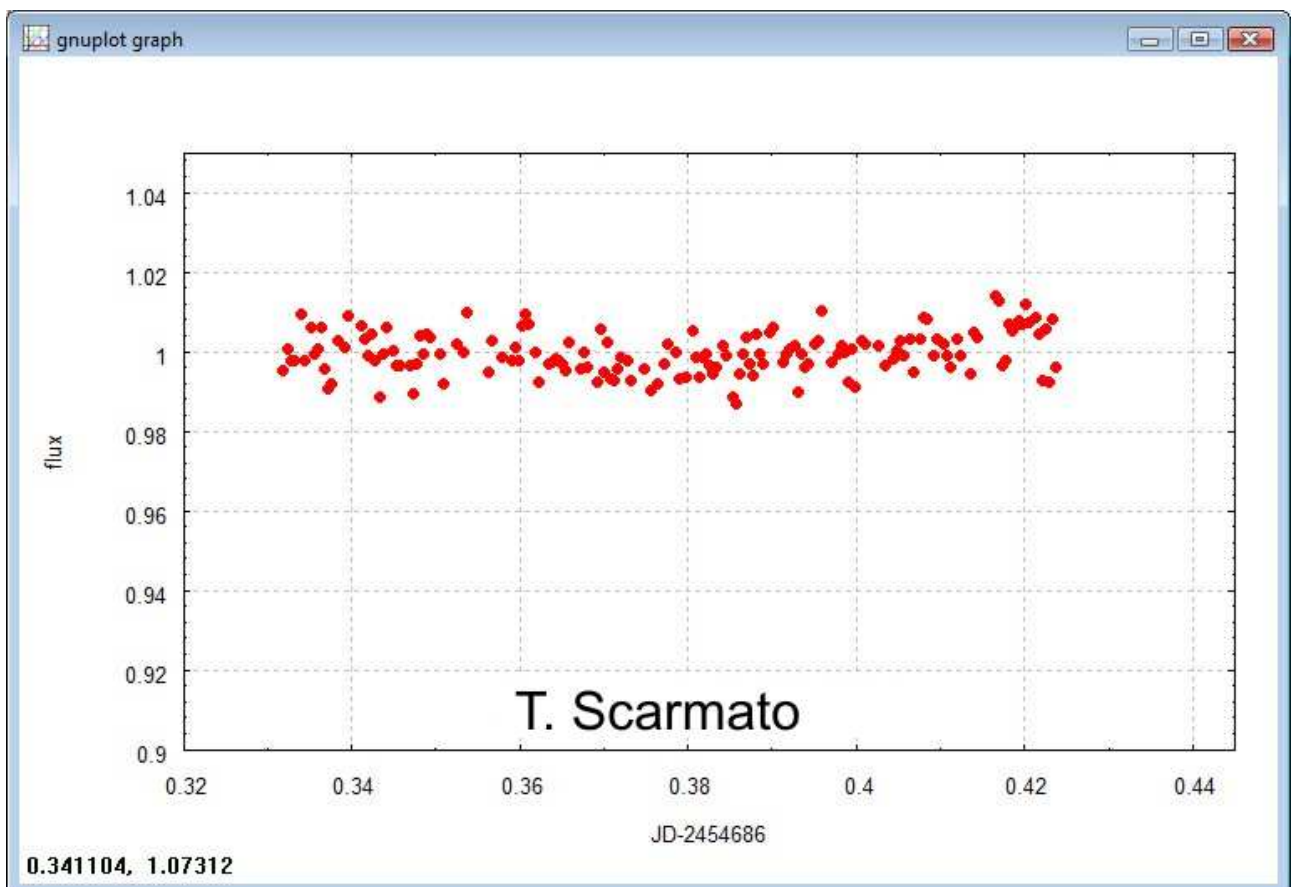

Fig. 1: The peak negative visible in the figure shows the change in magnitude of a star probably due to the transit of a planet in front of the star. The depth is about 8 thousandths of magnitude with an rms of 0.005 magnitudes. The shots were made with a Newton aperture 250 mm, focal length 1200 mm, Atik 16IC CCD camera and filter Rc photometric (for characteristics see table 2).

To make differential photometry is necessary to collect as many photons from the sources and then compare the resulting figures. As the number of photons that gather for a given source crucially depends on its brightness, if a source emits a steady stream, the number of photons that arrive at our detector will be settled within experimental error due to various causes, such as variation environmental conditions, temperature, etc … If instead a source does not emit a steady flux, the number of photons that arrive will vary over time. If the change is small, for example in the order of thousandths of magnitude, then it can detect it is necessary to compare the flux of variable source with the stable flux of sources.

Let us therefore refer to a simple formula which calculates the relationship between the flow of the star in question and the total flow of other reference sources said. (See next par.)

Tab.1: CCD camera's characteristics

| Parameters | STARLIGHT EXPRESS SXL8 | ATIK 16IC |
|---|---|---|
| Area in Pixel array | 512X512 (262.144 pixels squares) | 659X494 (325,546 pixels squares) |
| Pixel size | 15x15 micron | 7.4x7.4 micron |
| Full well depth | 150.000e | 40.000e |
| Dark current | 1e per second a -30°C | <1e per second at -25°C |
| Peak spectral response | 530nM, 50% a 400 nM e 650 nM | 500 nM |
| Quantum efficiency | 30% a 530 nM | >50% a 500 nM |
| A-D converter | 12 bits | 16 bits |
| Readout noise | < 20e | 7e |
| Anti-blooming | Yes | yes |
| Cooling | Yes | yes |
| CCD Type | Philips FT12 | Sony ICX-424AL |
| CCD size (area sensitive) | 7.8x7.8 mm | 4,8x3,7 mm |

Tab.2: Characteristics of the Telescope and Filter

| Parameters | Telescope | | Parameters | Filter Rc |
|---|---|---|---|---|
| *Aperture* | *250 mm* | | *Productor* | *Schuler* |
| *Focal Lenght* | *1200* | | *Band* | *Larga* |
| *Scale with SXL8* | *2.57 arcsec/pixel* | | *Lambada peak* | *5978 A* |
| *Scale with Atik16IC* | *1.26 arcsec/pixel* | | *FWHM* | *1297 A* |
| *Optic* | *Newton* | | | |
| *Type* | *Reflector* | | | |
| *Fov with SXL8* | *22'x22'* | | | |
| *Fov with Atik16IC* | *14'x11'* | | | |

When calculating the scale relate to the formulas:

$$FOV(') = \frac{3428 * \dim(mm)}{focal(mm)}, \quad scale(arc\sec/pixel) = \frac{FOV}{\dim(pixels)} * 60$$

where dim is the effective size of the chip of CCD camera.

**PURPOSE OF THE RESEARCH**

As we have said before in recent times has been demonstrated that even an amateur can achieve scientifically significant photometric astronomical objects and in particular, in the detection of transits of extrasolar planets.

The technique is to resume telescope with CCD + filter R the star under investigation and field around it and then applying the differential photometry noted the decrease of magnitude of the star caused by the passage of the planet in front of it along the line of sight observers . In order to apply the differential photometry is necessary to have other stars in possibly the same brightness as that under consideration concerned. When the series of images, as long as possible, at least 3 or April 4 hours, the images are reduced by the calibration files, dark, flat and if necessary the bias. The images are then calibrated with a small program (IRIS) to measure the number of ADU of the stars in the field. When the measure for all the stars considered good according to their individual flows, the formula

$$Flux = \frac{AduStar}{\sum Adu(ref1, ref2.ref3....)}$$

This formula gives a relationship between the flow of the star under investigation and the total flow of the stars of reference. To normalize the values obtained, it is estimated the median and the relationship

$$\frac{FLUX(Star)}{Median}$$

it then constructs the graph using a linear function that provides the best fit for the data obtained.

If the star is magnitude <11, has been shown that the blurring of the telescope can increase the accuracy if the flow is sufficiently high crop that is at least equal to 70% of the dynamics of the CCD and less than 90%. The defocusing is a method that allows you to collect the same number of photons but spread over an area greater than the CCD camera. In this way you have a question on homogeneous flux and shrinking the scintillation which increases the error on the measure. The bigger the brightness of the star must be greater focus to get more precise. See results with transits.

For comets is not yet clear whether or not blur improves the accuracy of the measure. We will see a further example periodic comet on 6P who provided interesting results, however. Moreover, it is not a linear function used to rent the data, but is compared only flow from the comet with one of the stars of reference. For this we must choose the right time of exposure. In general, the exposure time must take into account the number of photons collected during a single exposure. For comets is not yet possible to establish a formula that allows them to calculate with good precision, and then takes into account that the higher the ratio S/N (Signal to Noise), is the best measure of the flow from the comet. The aim is to apply this technique to the study of comets photometric to show if you can reach the same levels of accuracy that are obtained in observing the transit and then see if the results are in agreement with those obtained using scientifically tested and accepted as valid. We then compared the results with comets of varying magnitude shoot with and / or without blurring. This will enable us to understand if you should decide whether or not the defocus of comets or whether it is better to opt for shooting without defocus.

## METHODOLOGY

The methodology is to compare the flow of the object under investigation with the flow of objects that are in the same field. Shooting as we said earlier should be made preferably with the filter for photometric R Johnson-Cousin or, for that comets can isolate the issue in the band in which the dust that can be found in the hair of the comet transmit light coming from the Sun. The series should be longer possible, at least two hours and exposure times to have an S / N highest possible but taking into account that the comet has moved in the individual recovery. If the telescope on equatorial mount must be well aligned with the pole and must be minimized problems bending and tracing. We stress however that the images with objects can be moved measured as the total flux is always captured by the pixels of the CCD.

It is important not include windows opening to be such as to contain the flow but not too large to incorporate any background stars near the source.

The images will then be calibrated with the dark, the flat and if necessary by the bias. If it is possible we must maintain stable temperature of the CCD by at least 0.5 degrees. If you are unsure of what is most important series of darks during the shooting in order to mediate all the images and obtain a master dark refers to all possible temperatures occurred throughout the recovery.

Depending on the brightness of the comet, the 'quantum efficiency of the CCD and taking into account that the R photometric filter reduces the total quantity of photons arriving in the CCD compared to shoot without a filter, unless you have a formula to get the appropriate time optimum exposure, is bypassing the problem of doing tests with increasing exposure times from at least 60 seconds. When the counting of ADU is equal to 70% of the level of saturation of the camera then you can already proceed to capture the full range of images. If you want a better S / N must be taken into account both its motion of the comet that does not exceed 90% of the level of saturation of the CCD.

Once the images are made to their development, first of all calibrating them then with a program that can measure the ADU of individual items automatically, you get the count for each image. In our case was used IRIS. The program, once measured the individual sources in the individual images, providing an output file with dates in JD - ADU and the values of the objects for each image records in chronological order. (see Table 3)

Tab. 3: sample records with the data, measure,
in the first column is the Julian Date observation on each image, in subsequent
columns are the counts the object that you want the photometry of stars and reference

| JD | FLUSSOJ1 | REF1 | REF2 |
|---|---|---|---|
| 2454718.3129500 | 21304 | 510508 | 44780 |
| 2454718.3136991 | 22548 | 521433 | 45027 |
| 2454718.3144502 | 22145 | 518231 | 39568 |
| 2454718.3152014 | 23422 | 515335 | 42985 |
| 2454718.3159549 | 22926 | 509435 | 42423 |
| 2454718.3167083 | 22180 | 512611 | 37962 |
| 2454718.3174618 | 25370 | 503666 | 43654 |
| 2454718.3182141 | 25005 | 512055 | 37463 |
| 2454718.3189653 | 21340 | 504458 | 38060 |
| 2454718.3197176 | 24238 | 499754 | 41179 |
| 2454718.3204699 | 23582 | 513459 | 46686 |
| 2454718.3212222 | 23654 | 510294 | 39781 |
| 2454718.3219745 | 21456 | 502620 | 38062 |
| 2454718.3227269 | 25412 | 507747 | 38987 |
| 2454718.3234792 | 23214 | 499630 | 35155 |
| 2454718.3242315 | 20154 | 498266 | 38123 |
| 2454718.3249838 | 20619 | 510332 | 37554 |
| 2454718.3257361 | 24418 | 493006 | 38936 |
| 2454718.3264884 | 22754 | 496654 | 40879 |
| 2454718.3272407 | 21454 | 493840 | 37073 |

Because the program can do so automatically photometry on only 5 items at a time, if you need more than 5 photometric objects, you get more than one file with the data of individual objects.

To put them together you can use Excel and then create a single file type. Txt to be able to develop a program of analysis of the curves of light. In our case was used GNUPlot. It then proceeds with the comparison between the flux of the object under investigation and the flux of the other stars measured in the field. The formula described above

$$Flux = \frac{AduStar}{\sum Adu(ref1, ref2.ref3....)}$$

permit to obtain the values that are fitted by a linear function that determines the best performance of our points of the curve of light. The sum in the denominator to obtain a theoretical star whose flow is equal to the sum of the flows of individual stars. In this way you can build a chart that if precision is desired, highlights the brightness variations that appear as positive or negative peaks in the curve of light. (see first results). As for the comets is not yet clear whether or not blur helps to increase the accuracy of the measures also bright comet, so this research will tell us what are the limits of the method of photometry differential applies to comets. We will see in the next paragraph the results and then we will draw conclusions based upon qualitative and quantitative them.

# RESULTS AND DISCUSSION

## *Photometry of the Comet 73P/Schwassmann-Wachmann B*

Here are the orbital elements of the component b of the comet 73/P.

```
Epoch 2006 May 25.0 TT = JDT 2453880.5
T 2006 June 7.92466 TT              MPC
q  0.9390718     (2000.0)    P           Q
n  0.18394396   Peri. 198.80779   -0,02873648   +0.98221260
a  3.0620506    Node  69.88766   -0,88989677   +0.05940885
e  0.6933193    Incl. 11.39726   -0,45525592   -0,17812634
P  5.36
```

    Comet 73/P is a periodical that has fragmented several times in its repeated passes at perihelion. The various pieces are away from each other because of the initial thrust due to the process of fragmentation that still has not been explained fully.

    The brighter components, c and b, during the last passage to perihelion showed strong activity with further fragmentation and an interesting photometric evolution.

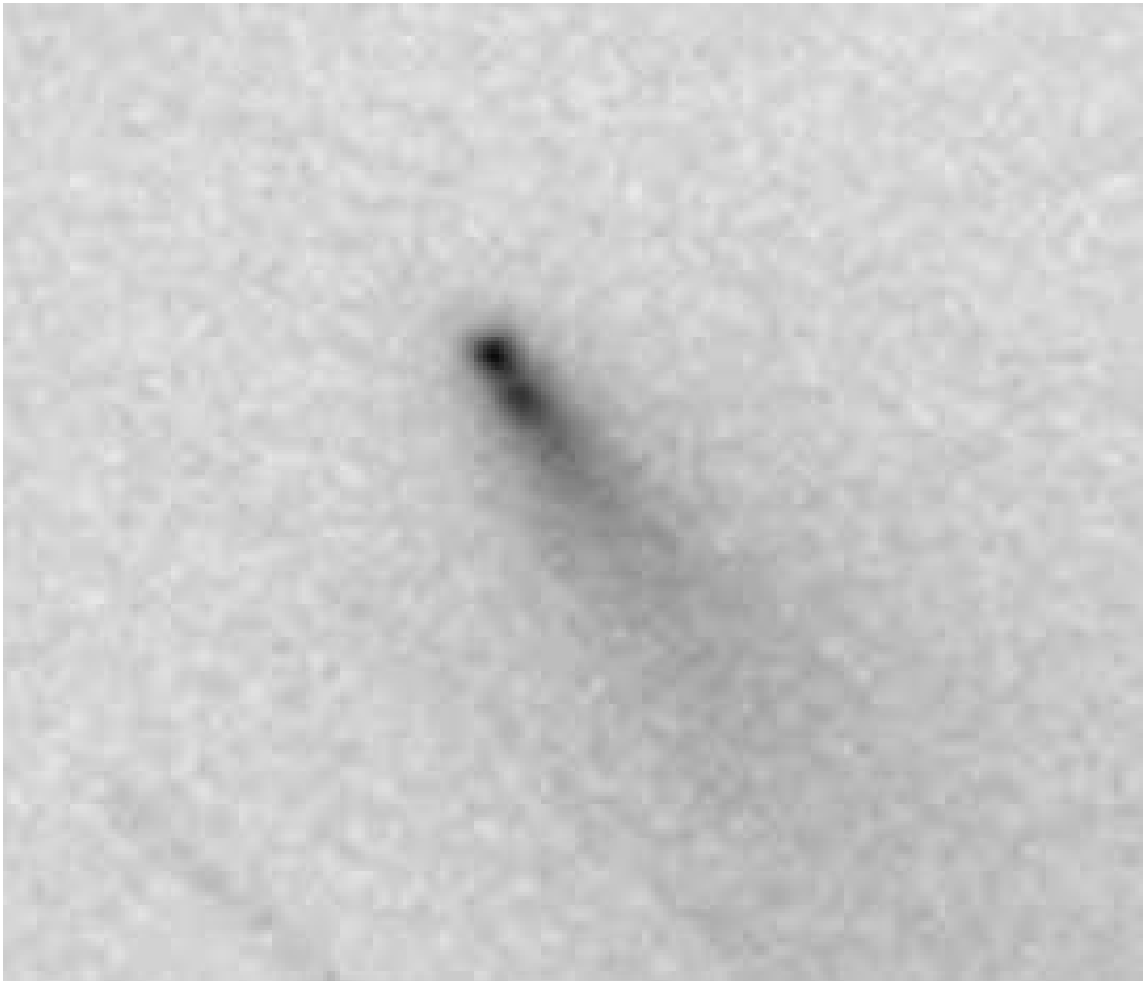

Fig.2: Comet 73/Pb imaged the evening of April 22, 2006

In the first decade of April the comet 73/P-b had an outburst resulting in fragmentation of the nucleus. In the evenings from April 20-24 was possible to see the piece that was seconded by the core and is slowly away towards South-West. (see Fig.2) The repeated outburst and fragmentation have been registered by many observers.
(See link http://digilander.libero.it/infosis/homepage/astronomia/73p.htm)

For our analysis we chose the images of 28 April 2006 of component b was apparently quiet after a strong outburst occurred in the previous week. The series of images covering a time interval of about 2 hours.

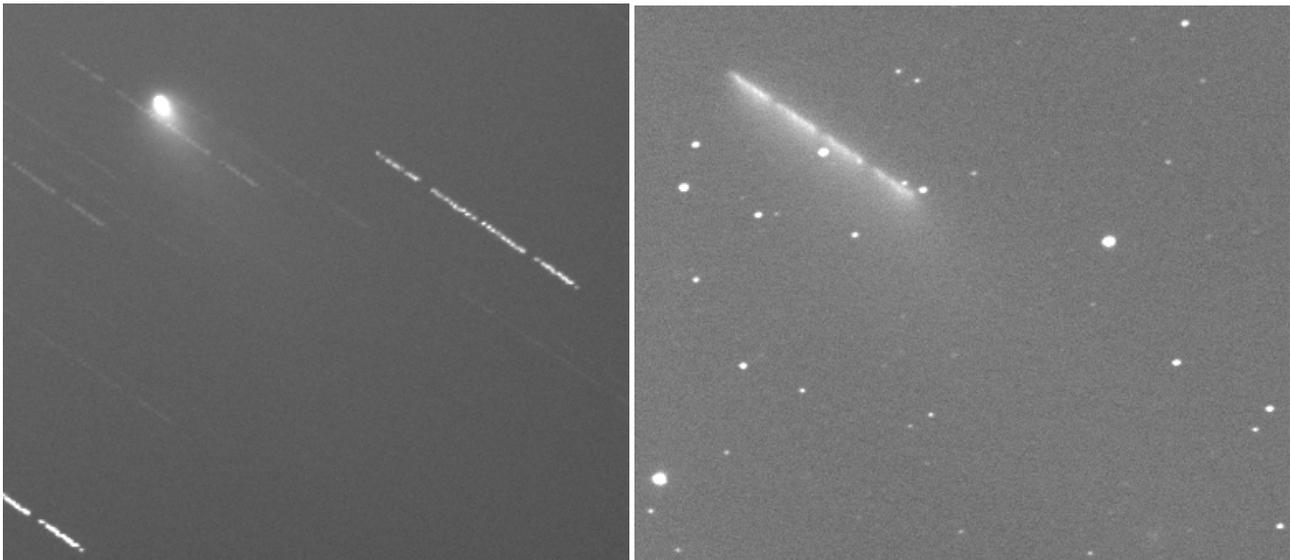

Fig. 3: The comet 73/Pb resumed the evening of 28 April 2006. On the left image was obtained with mediating Astroart 150 frames for 30 sec each, and aligning on the photometric centre of the comet. On the right the same images have been mediated by aligning on the centre of photometric stars. The error of alignment is 1-2 pixels. Note that near the comet there are stars in the field who do not have permission to use openings more than 9 pixels as the counts were altered as a result of their presence.

The comet was at 0.147 U. A. from Earth and the heliocentric distance was 1104 U. A. Because of its proximity to Earth on its own motion of the comet was high, you could appreciate in a few minutes observing with a 7x50 binoculars. For this reason, the exposures were only 30 seconds each. The shots were made Newtonian telescope with a 250 mm aperture, focal length 1200 mm, with a CCD camera Starlight Express SXL8 Rc filter and direct fire. (for the complete system see Tables 1 and 2). As we said in the introduction, for our purposes it is necessary that the comet is that the stars of reference remain the same field throughout the recovery. Consequently only two stars were adapted from a photometric to perform with good precision photometry of the comet(see Table 4)

| *Star/Catalog* | mag. - USNO B1 | mag. - TYCHO 2 | mag. - SDSS |
|---|---|---|---|
| T 2574:136:1 (Ref1) | R1=8.95 – R2=8.89 | Bt=10.946 – Vt=10,629 | R=9.340 |
| T 2574:578:1 (Ref2) | R1=9.77 – R2=9.69 | Bt=12.147 – Vt=10,629 | R=10.119 |
| | | | |

Tab. 4: photometric parameters of the stars of reference derived from catalogues on-line using the Aladin, after calibrated astrometry the image aligned on the stars of the previous figure.

The photometric results obtained can be seen in the figures below.

|  | Km/UA | In meters |
|---|---|---|
| **Rho (Coma Radius)** | 2500 | 2500000 |
| *D (Earth-Comet)* | 0,1470 | 22050000000 |
| R (Sun-Comet) | 1,1040 | 165133008 |

Aperture 9 pixels, Resolution 2.57 arcsec/pixel 25 cm Telescope Newton f/4.8 + CCD Starlight Express SXL8 + Rc

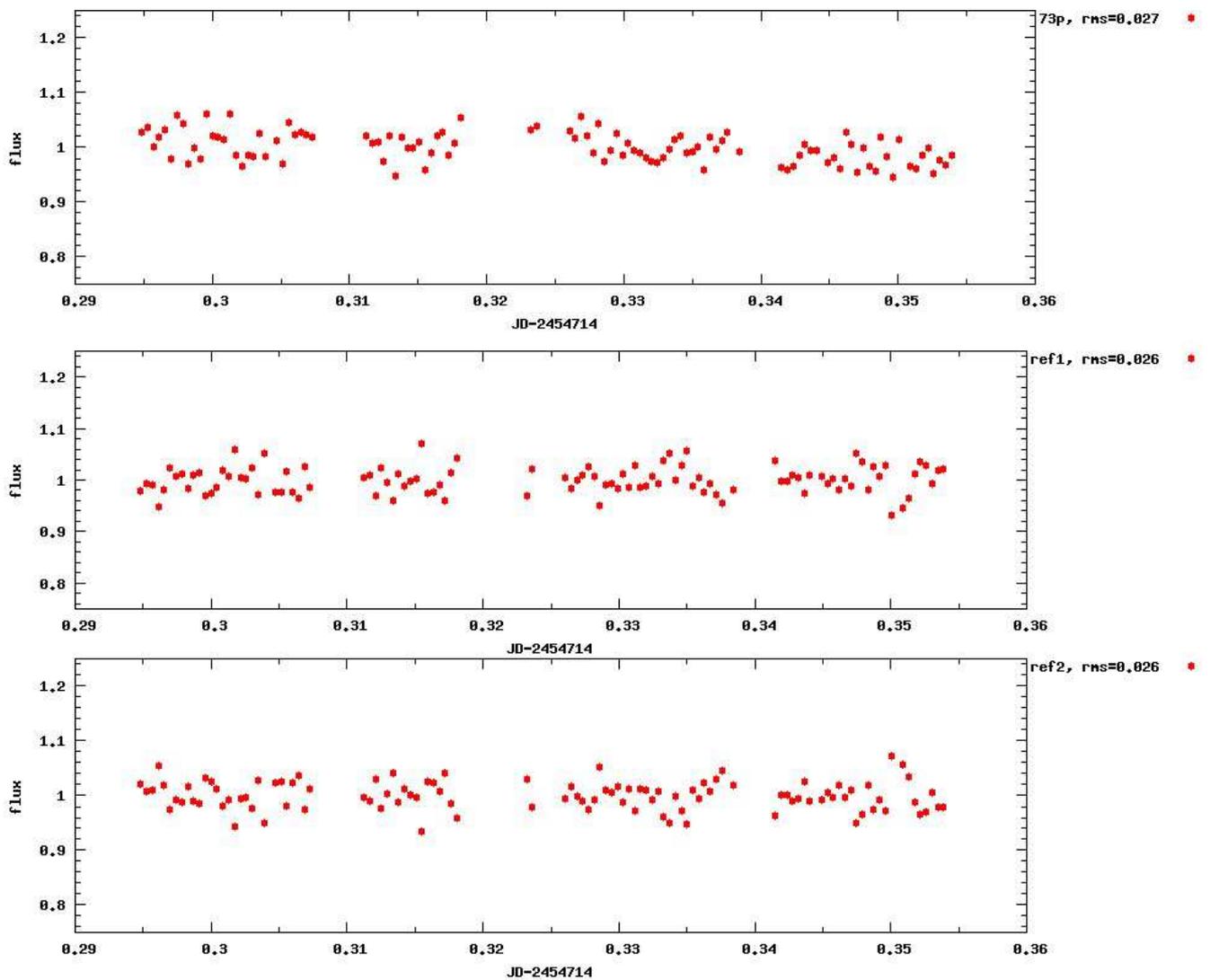

Fig. 4: The graphs show the flux of the comet 73/P component b compared with the flux of the stars of reference (see Table 4 for the characteristics of the stars). The rms respectively 0.027 for comet and 0.026 for the two stars of reference. You can notice a decrease in the flux of the comet after JD = 0.33.

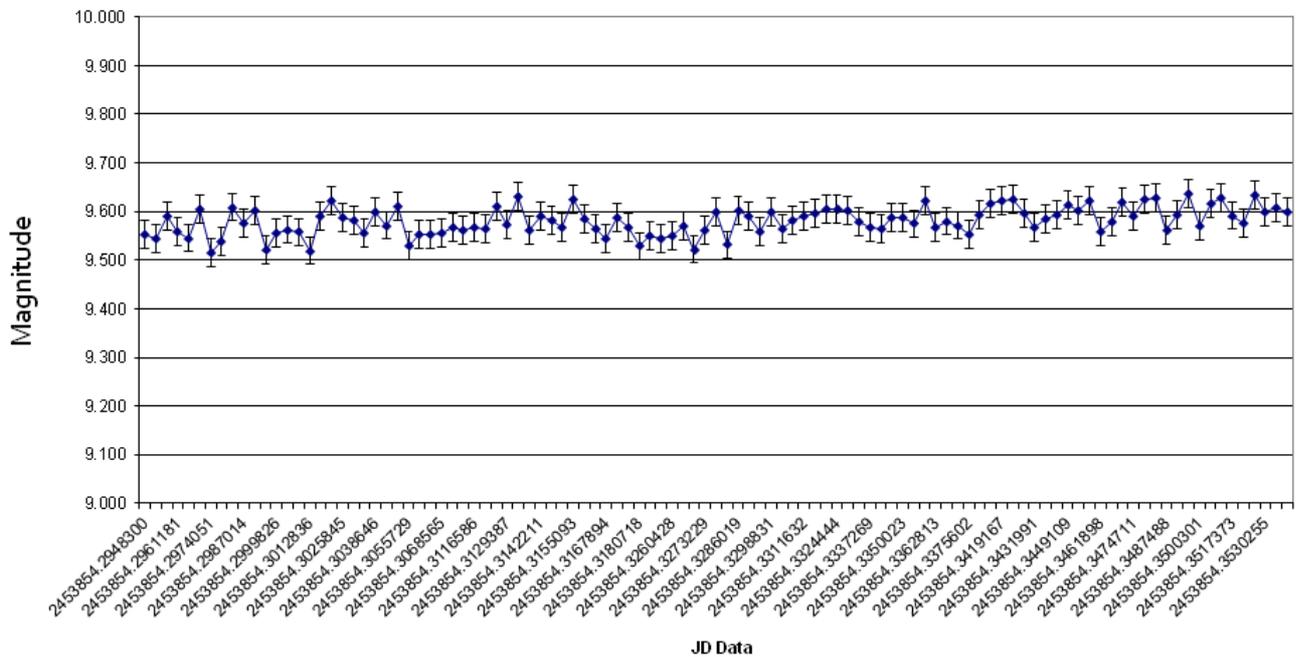

Fig. 5: The graph of the magnitude of the comet 73/P with the flux converted into magnitudes, showing an "increase" after JD = 0.33, about 0.1 magnitudes with an rms of 0.026.

Table 5 contains all of the counts of comet and star of reference for the complete set of images.

Tab. 5: fluxes of comet 73/P and stars measured in reference ADU.

| | | | | |
|---|---|---|---|---|
| 2453854.2948300 | 43122 | 77148 | 42716 | 7462 |
| 2453854.2952639 | 43263 | 77186 | 42118 | 7466 |
| 2453854.2956910 | 41906 | 77440 | 42334 | 8245 |
| 2453854.2961181 | 43726 | 77987 | 44574 | 7256 |
| 2453854.2965451 | 43511 | 77571 | 42832 | 7124 |
| 2453854.2969722 | 42395 | 80968 | 42784 | 7360 |
| 2453854.2974051 | 46635 | 81939 | 44015 | 6851 |
| 2453854.2978368 | 44829 | 79937 | 42819 | 7766 |
| 2453854.2982697 | 42082 | 79934 | 44021 | 6835 |
| 2453854.2987014 | 42543 | 79241 | 42499 | 6581 |
| 2453854.2991285 | 39404 | 75079 | 40089 | 6559 |
| 2453854.2995556 | 41024 | 70964 | 39651 | 6960 |
| 2453854.2999826 | 42848 | 77175 | 42878 | 6795 |
| 2453854.3004178 | 44315 | 80350 | 44099 | 7523 |
| 2453854.3008507 | 41299 | 75986 | 40397 | 5986 |
| 2453854.3012836 | 41706 | 73070 | 39240 | 7017 |
| 2453854.3017153 | 37008 | 71079 | 36351 | 5587 |
| 2453854.3021516 | 32503 | 62588 | 33743 | 5797 |
| 2453854.3025845 | 32622 | 61504 | 33217 | 4651 |
| 2453854.3030116 | 29289 | 55679 | 29462 | 3622 |
| 2453854.3034387 | 34160 | 61201 | 34083 | 5869 |
| 2453854.3038646 | 32757 | 62828 | 32372 | 5725 |
| 2453854.3047188 | 38450 | 69845 | 38685 | 6905 |
| 2453854.3051458 | 38795 | 73453 | 40774 | 6460 |
| 2453854.3055729 | 41520 | 74084 | 39418 | 6367 |
| 2453854.3060012 | 42677 | 76695 | 42539 | 6723 |
| 2453854.3064294 | 44801 | 79864 | 44848 | 7553 |
| 2453854.3068565 | 44750 | 81793 | 43188 | 7384 |
| 2453854.3072824 | 46080 | 83451 | 45802 | 8598 |
| 2453854.3112338 | 42389 | 77093 | 41590 | 7351 |
| 2453854.3116586 | 44796 | 82719 | 44337 | 7018 |
| 2453854.3120868 | 43952 | 79834 | 44569 | 6690 |
| 2453854.3125127 | 39393 | 75638 | 40026 | 6990 |
| 2453854.3129387 | 39414 | 71519 | 38906 | 7971 |
| 2453854.3133692 | 37629 | 72553 | 40967 | 5745 |
| 2453854.3137951 | 48091 | 87991 | 47044 | 7859 |
| 2453854.3142211 | 38100 | 70384 | 38582 | 7344 |
| 2453854.3146539 | 34730 | 64462 | 34974 | 5771 |
| 2453854.3150810 | 32800 | 60279 | 32529 | 5181 |
| 2453854.3155093 | 28434 | 56282 | 28488 | 4946 |
| 2453854.3159363 | 20492 | 38034 | 21139 | 3135 |
| 2453854.3163634 | 38112 | 68684 | 38106 | 6860 |
| 2453854.3167894 | 41049 | 73896 | 40353 | 5826 |
| 2453854.3172164 | 41467 | 76812 | 43358 | 6273 |
| 2453854.3176447 | 48147 | 89003 | 47486 | 7535 |
| 2453854.3180718 | 48149 | 85850 | 44553 | 8509 |
| 2453854.3232025 | 42562 | 75731 | 42256 | 7224 |
| 2453854.3236319 | 45075 | 81118 | 43025 | 7709 |
| 2453854.3260428 | 49174 | 88664 | 47766 | 8202 |
| 2453854.3264699 | 50100 | 90785 | 49995 | 9423 |
| 2453854.3268981 | 52865 | 92734 | 50150 | 8794 |
| 2453854.3273229 | 49881 | 90974 | 48768 | 8567 |
| 2453854.3277500 | 48863 | 92440 | 48795 | 9181 |
| 2453854.3281748 | 47856 | 85336 | 45869 | 7236 |
| 2453854.3286019 | 45779 | 85540 | 48769 | 6927 |
| 2453854.3290289 | 48186 | 89477 | 48941 | 8705 |
| 2453854.3294560 | 48403 | 87292 | 47545 | 8631 |
| 2453854.3298831 | 47652 | 89083 | 49035 | 8768 |
| 2453854.3303102 | 48495 | 89567 | 47982 | 6933 |
| 2453854.3307373 | 47780 | 88657 | 48642 | 7512 |
| 2453854.3311632 | 49677 | 94014 | 49555 | 8356 |
| 2453854.3315891 | 48846 | 91980 | 50490 | 7354 |
| 2453854.3320174 | 48170 | 91270 | 49957 | 8058 |
| 2453854.3324444 | 48760 | 93201 | 50053 | 7784 |
| 2453854.3328727 | 49710 | 93820 | 51172 | 8330 |
| 2453854.3333009 | 49407 | 93070 | 48513 | 7796 |
| 2453854.3337269 | 36239 | 67466 | 34721 | 6233 |
| 2453854.3341516 | 47106 | 85642 | 46316 | 8508 |
| 2453854.3345764 | 50107 | 94855 | 49986 | 7660 |
| 2453854.3350023 | 47526 | 90568 | 46456 | 7834 |
| 2453854.3354271 | 43731 | 80639 | 44166 | 6850 |
| 2453854.3358542 | 45258 | 87600 | 47171 | 7502 |
| 2453854.3362813 | 45449 | 81987 | 45485 | 8433 |
| 2453854.3367083 | 48856 | 90578 | 49431 | 7848 |
| 2453854.3371331 | 48748 | 88361 | 49286 | 8700 |
| 2453854.3375602 | 47910 | 85082 | 48206 | 8212 |
| 2453854.3384132 | 51688 | 95900 | 52899 | 9387 |
| 2453854.3414873 | 51208 | 99782 | 52024 | 8618 |
| 2453854.3419167 | 52320 | 101028 | 54772 | 8960 |
| 2453854.3423437 | 52349 | 100523 | 54490 | 9896 |
| 2453854.3427708 | 53073 | 100132 | 53709 | 9007 |
| 2453854.3431991 | 53526 | 98937 | 53319 | 7907 |
| 2453854.3436273 | 53347 | 98506 | 54782 | 8884 |
| 2453854.3440556 | 53216 | 99591 | 53443 | 9803 |
| 2453854.3449109 | 52374 | 100087 | 53787 | 9624 |
| 2453854.3453368 | 52767 | 99574 | 54305 | 8875 |
| 2453854.3457616 | 51486 | 99416 | 53650 | 8925 |
| 2453854.3461898 | 54220 | 97200 | 53627 | 10159 |
| 2453854.3466169 | 53521 | 98869 | 53360 | 9400 |
| 2453854.3470440 | 50826 | 98262 | 53814 | 7222 |
| 2453854.3474711 | 52060 | 98320 | 50659 | 10029 |
| 2453854.3478981 | 51398 | 100018 | 52322 | 9785 |
| 2453854.3483229 | 51111 | 98508 | 54326 | 8762 |
| 2453854.3487488 | 53656 | 98440 | 51956 | 9259 |
| 2453854.3491759 | 52244 | 98781 | 53077 | 8182 |
| 2453854.3496030 | 51213 | 101466 | 53432 | 8231 |
| 2453854.3500301 | 50098 | 89265 | 51892 | 8906 |
| 2453854.3508843 | 51537 | 97129 | 55581 | 8439 |
| 2453854.3513113 | 50280 | 95883 | 53773 | 9410 |
| 2453854.3517373 | 53831 | 101658 | 54352 | 8641 |
| 2453854.3521678 | 53745 | 100975 | 52800 | 8076 |
| 2453854.3525984 | 50911 | 100137 | 52666 | 9049 |
| 2453854.3530255 | 51764 | 98159 | 53469 | 7809 |
| 2453854.3534525 | 52435 | 101186 | 53714 | 8057 |
| 2453854.3538785 | 52690 | 99882 | 52924 | 9468 |

Using formula,

$$\text{magstar} = -2.5 * \text{LOG}((1 + (\text{aduref2} + \text{aduref3}) / \text{aduref1})) + \text{magref1},$$

where magref1 R is the magnitude of the star brighter used as a reference, obtained from USNOB1 catalogue, (see Table 4), you get the magnitude of the star theoretical reference.

Through the formula

$$\text{magcomet} = -2.5 * \text{LOG}(\text{aducomet}/(\text{aduref1} + \text{aduref2} + \text{aduref3})) + \text{magstar}$$

you have the magnitude of the comet. The median values derived with the fluxes of Table 5 are:

**magstar = 8.378 (+/-0.012), magcomet=9.581(+/-0.029).**

## Photometry of the Comet C/2008 J1 (Boattini)

Here are the orbital elements of the comet C/2008 J1

```
C/2008 J1 (Boattini)
Epoch 2008 Nov. 30.0 TT = JDT 2454800.5
T 2008 July 13.2757 TT                MPC
q   1.724289      (2000.0)       P         Q
z  +0.006102    Peri.  68.1304    +0.4602796    +0.1204894
+/- 0.000017    Node  273.4188   -0.6424076    +0.7290203
e   0.989478    Incl.  61.7787    +0.6127440    +0.6738039
```

      Comet C/2008 J1 discovery from the Italian astronomer A. Boattini, had a seemingly stable photometric performance but showed a very peculiar shape-behaviour. (See fig. 6)

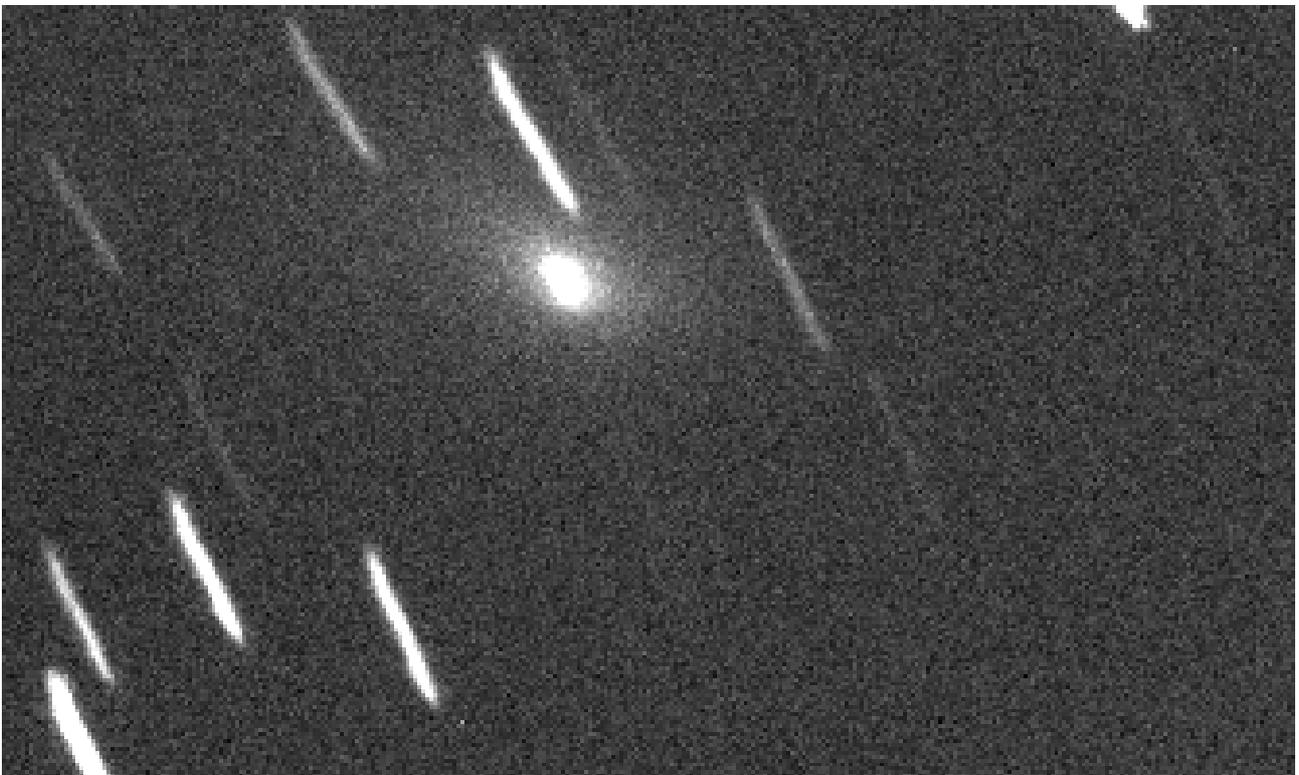

Fig. 6: Comet C/2008 J1 taken on 4 September 2008 with a Newton Telescope 250 mm, focal length 1200 mm CCD camera and filter Rc.

As you can see the coma is very asymmetrical with a part that stretches to the north-east and the other almost in opposition in south-west. This form has remained stable throughout the period of our observations. The series of images for our research were obtained on the evening on 4 September 2008, when the comet was in position circumpolar. There were obtained 120 images from 30 seconds each with a telescope Newton 250 mm aperture, focal length 1200 mm, Atik CCD filter and 16IC Rc direct fire. (for the complete system see Tables 1 and 2). The pursuit of hype was not optimal for different images are moved, but were the same used for granting the photometry. The stars photometric shown good are 4 and listed in Table 6.

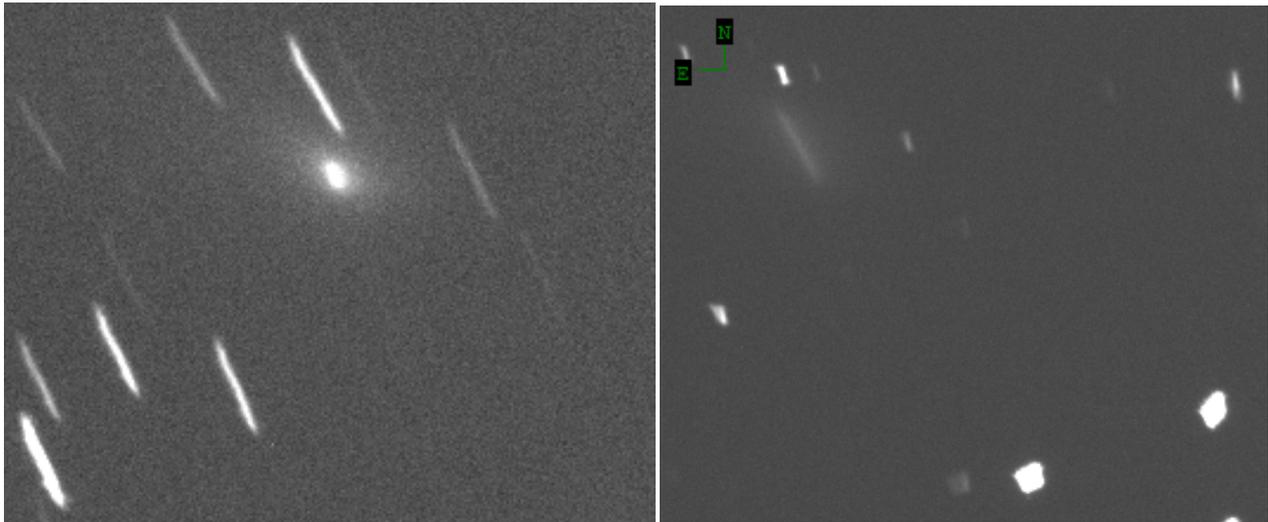

Fig. 7: Comet C/2008 J1 observed the evening on 4 September 2008. On the left image was obtained with mediating Astroart 120 frames for 30 sec each, and aligning on the photometric centre of the comet. On the right the same images have been mediated by aligning on the centre of photometric stars. The error of alignment is 2-3 pixels with the stars moved because of bad pursuit.

| *Stella/Catalogo* | mag. - USNO B1 | mag. - TYCHO 2 | mag. – GSC 2.2 |
|---|---|---|---|
| T 4637:1159:1 (Ref1) | R1=11.27– R2=11.23 | Bt=12.349 – Vt=11.682 | |
| T 4637:617:1(Ref2) | R1=11.13 – R2=11.06 | Bt=13.956 – Vt=12.105 | |
| GSC1393 (Ref3) | R1=11.32 - R2=11.15 | | R=11.63 |
| GSC1353 (Ref4) | R1=11.32 - R2=11.38 | | R=11.88 |
| | | | |

Tab. 6: photometric parameters of the stars of reference derived from catalogs on-line using Aladin, after calibrated astrometry the image aligned on the stars of the previous figure.

The photometric results obtained can be seen in the figures below.

*Comet aperture 9 pixels - Resolution 1.26 arcsec / pixels - 25 cm Ttelescope Newton f / 4.8 + CCD Atik 16IC + Rc*

| | Km/UA | In Meters |
|---|---|---|
| **Rho (Coma Radius)** | 8086 | 8086000 |
| *D (Earth-Comet)* | 1,7480 | 2,622E+11 |
| **R (Sun-Comet)** | 1,8590 | 278063643 |

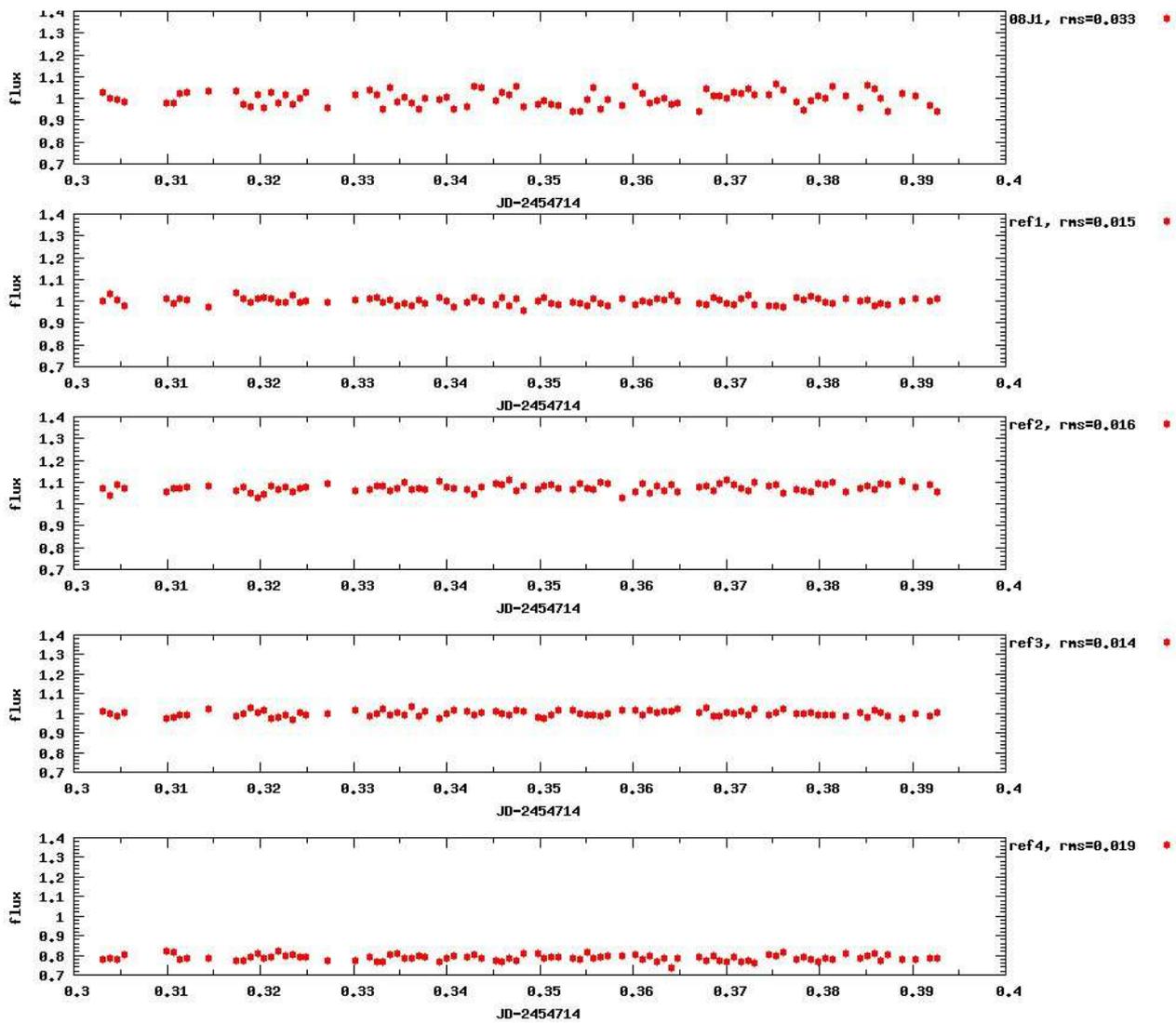

Fig. 8: The graphs show the flux of Comet C/2008 J1 compared with the flux of the stars of reference (see table 6 to the characteristics of the stars). The rms respectively 0.033 for the comet and 0.015, 0.016, 0.014 and 0.019 for the four stars of reference. In this case you can see a steady flux to the comet

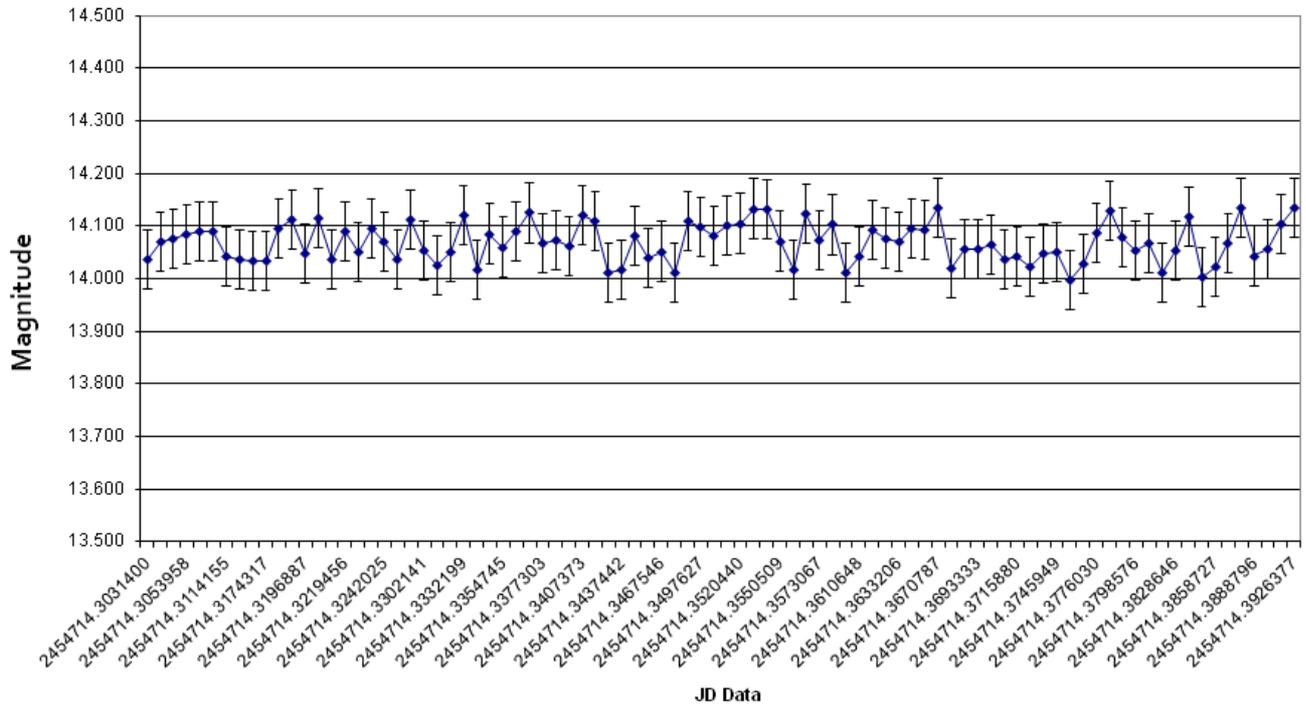

Fig. 9: The graph of the magnitude of Comet C/2008 J1 with the flux converted into magnitudes.

Following table contains all of the counts of comet and stars of reference for the complete set of images.

Tab. 7: flux of comet C/2008 j1 and reference stars measured in ADU.

| JD | FLUX J1 | REF1 | REF2 | REF3 | REF4 |
|---|---|---|---|---|---|
| 2454714.3031400 | 18874 | 245875 | 222713 | 212977 | 173819 |
| 2454714.3038912 | 18420 | 253472 | 218828 | 212601 | 176458 |
| 2454714.3046435 | 18142 | 245895 | 224638 | 208284 | 173900 |
| 2454714.3053958 | 18133 | 242647 | 223286 | 213281 | 179610 |
| 2454714.3099109 | 17733 | 244434 | 216729 | 203897 | 178731 |
| 2454714.3106632 | 17668 | 240236 | 219056 | 204217 | 177629 |
| 2454714.3114155 | 18245 | 240671 | 216362 | 204213 | 169502 |
| 2454714.3121678 | 18669 | 244127 | 221165 | 207542 | 173659 |
| 2454714.3144248 | 18442 | 235001 | 218930 | 209665 | 170851 |
| 2454714.3174317 | 18405 | 245570 | 214685 | 203178 | 168355 |
| 2454714.3181840 | 17489 | 242733 | 218669 | 206881 | 169021 |
| 2454714.3189363 | 17133 | 238523 | 213165 | 210145 | 171222 |
| 2454714.3196887 | 18018 | 239488 | 208716 | 204651 | 173325 |
| 2454714.3204410 | 17084 | 241981 | 212967 | 208017 | 170167 |
| 2454714.3211933 | 18301 | 240135 | 217480 | 201351 | 171155 |
| 2454714.3219456 | 17821 | 242549 | 219252 | 206058 | 179801 |
| 2454714.3226979 | 18059 | 237805 | 216615 | 203213 | 171788 |
| 2454714.3234502 | 17410 | 244399 | 213869 | 201020 | 173868 |
| 2454714.3242025 | 17933 | 239360 | 218067 | 207680 | 172780 |
| 2454714.3249537 | 18619 | 242001 | 220301 | 206994 | 174190 |
| 2454714.3272060 | 17375 | 241884 | 223142 | 208855 | 171438 |
| 2454714.3302141 | 18157 | 241299 | 215738 | 209070 | 169150 |
| 2454714.3317164 | 18689 | 242728 | 217818 | 205311 | 173234 |
| 2454714.3324688 | 18278 | 244125 | 219442 | 206704 | 168348 |
| 2454714.3332199 | 17000 | 238077 | 217973 | 209619 | 167213 |
| 2454714.3339722 | 18904 | 242767 | 217060 | 206526 | 176056 |
| 2454714.3347222 | 17479 | 234015 | 215369 | 205403 | 174140 |
| 2454714.3354745 | 17880 | 236483 | 219168 | 203051 | 169807 |
| 2454714.3362269 | 17294 | 233079 | 213202 | 208774 | 168975 |
| 2454714.3369792 | 16814 | 238769 | 214999 | 201713 | 171643 |
| 2454714.3377303 | 17553 | 233497 | 212183 | 203835 | 168614 |
| 2454714.3392338 | 17482 | 238611 | 217952 | 199011 | 164718 |
| 2454714.3399861 | 17598 | 234458 | 213103 | 201071 | 167594 |
| 2454714.3407373 | 16847 | 232583 | 214739 | 206655 | 171227 |
| 2454714.3422407 | 16874 | 234063 | 211729 | 203823 | 168908 |
| 2454714.3429919 | 18457 | 237978 | 208213 | 200498 | 170823 |
| 2454714.3437442 | 18464 | 236291 | 214609 | 203953 | 168025 |
| 2454714.3452477 | 17493 | 235147 | 218660 | 206231 | 166720 |
| 2454714.3460000 | 18119 | 239310 | 215962 | 202415 | 164812 |
| 2454714.3467546 | 17875 | 232375 | 219524 | 202079 | 167912 |
| 2454714.3475069 | 18659 | 239577 | 213984 | 206503 | 167367 |
| 2454714.3482593 | 17267 | 233063 | 219596 | 208560 | 175925 |
| 2454714.3497627 | 17171 | 236689 | 213270 | 200364 | 172910 |
| 2454714.3505139 | 17285 | 237687 | 214304 | 197557 | 167595 |
| 2454714.3512662 | 16870 | 231404 | 213348 | 198920 | 167256 |
| 2454714.3520440 | 16721 | 229285 | 210145 | 201836 | 166196 |
| 2454714.3535463 | 16252 | 230441 | 208440 | 201572 | 164855 |
| 2454714.3542986 | 16436 | 232161 | 215050 | 201111 | 165823 |
| 2454714.3550509 | 17394 | 230301 | 212043 | 199631 | 172361 |
| 2454714.3558032 | 18354 | 236899 | 212054 | 201038 | 167810 |
| 2454714.3565544 | 16942 | 237107 | 219957 | 203007 | 171447 |
| 2454714.3573067 | 17700 | 234630 | 218623 | 204776 | 171500 |
| 2454714.3588102 | 17339 | 242049 | 211012 | 209180 | 173012 |
| 2454714.3603137 | 18890 | 237448 | 215079 | 209571 | 174886 |
| 2454714.3610648 | 18269 | 238738 | 219464 | 204326 | 169609 |
| 2454714.3618171 | 17571 | 239704 | 215044 | 209232 | 173936 |
| 2454714.3625694 | 17439 | 237467 | 214809 | 203765 | 164596 |
| 2454714.3633206 | 17779 | 239445 | 214869 | 207153 | 169952 |
| 2454714.3640718 | 17362 | 243125 | 218798 | 207177 | 161879 |
| 2454714.3648241 | 17622 | 241400 | 215737 | 211213 | 172211 |
| 2454714.3670787 | 16919 | 239450 | 218923 | 208054 | 172940 |
| 2454714.3678287 | 18955 | 239932 | 221623 | 212892 | 171010 |
| 2454714.3685810 | 18255 | 244619 | 217409 | 205596 | 175083 |
| 2454714.3693333 | 18072 | 240864 | 220085 | 204083 | 168337 |
| 2454714.3700845 | 18257 | 241872 | 226881 | 210498 | 170606 |
| 2454714.3708356 | 18525 | 238506 | 221179 | 207278 | 172561 |
| 2454714.3715880 | 18640 | 246568 | 221219 | 211742 | 170405 |
| 2454714.3723391 | 18594 | 243821 | 215176 | 204609 | 168390 |
| 2454714.3730914 | 18377 | 239463 | 223369 | 212002 | 167849 |
| 2454714.3745949 | 18127 | 236085 | 217971 | 205026 | 174093 |
| 2454714.3753484 | 19155 | 236740 | 219834 | 206834 | 173195 |
| 2454714.3760995 | 18429 | 233979 | 212231 | 208493 | 174678 |
| 2454714.3776030 | 17626 | 243636 | 217492 | 206478 | 170289 |
| 2454714.3783542 | 16972 | 241982 | 216403 | 207006 | 172598 |
| 2454714.3791053 | 17726 | 243870 | 214632 | 206912 | 170358 |
| 2454714.3798576 | 18090 | 241292 | 219561 | 204245 | 167541 |
| 2454714.3806088 | 17750 | 236792 | 217340 | 203519 | 169424 |
| 2454714.3813611 | 18682 | 236327 | 219197 | 203524 | 168310 |
| 2454714.3828646 | 17944 | 239353 | 212514 | 201802 | 173039 |
| 2454714.3843692 | 16800 | 235332 | 213562 | 203659 | 167766 |
| 2454714.3851215 | 18727 | 237857 | 215517 | 199941 | 170392 |
| 2454714.3858727 | 18076 | 228357 | 209310 | 201855 | 169326 |
| 2454714.3866250 | 17576 | 234272 | 216829 | 203240 | 166036 |
| 2454714.3873762 | 16704 | 235636 | 218036 | 203021 | 172795 |
| 2454714.3888796 | 17942 | 235660 | 218047 | 198438 | 166511 |
| 2454714.3903831 | 17678 | 235787 | 212758 | 201402 | 165550 |
| 2454714.3918866 | 17043 | 236001 | 216339 | 201378 | 168996 |
| 2454714.3926377 | 16564 | 238188 | 211582 | 203442 | 168518 |

The median values derived with the flux of Table 7 are the following:

**magstar=9.895(+/-0.012), magcomet=14.068 (+/-0.036**)

## Photometry of the Comet 6P/d'Arrest

Here are the orbital elements of the comet 6P

```
6P/d'Arrest
Epoch 2008 Aug. 2.0 TT = JDT 2454680.5
T 2008 Aug. 14.9589 TT              MPC
q   1.353507       (2000.0)     P          Q
n   0.1507993   Peri.  178.1196   +0.7332504   +0.6435761
a   3.495718    Node   138.9358   -0.6281291   +0.7646977
e   0.612810    Incl.   19.5148   -0.2603801   -0.0323621
P   6.54
```

The periodic 6P, with a period of 6.54 years, has repeatedly made the halfway point and has been studied in depth. The last step to the minimum distance from the sun happened on 2 August 2008 and was seen almost daily for a discrete period of time. It was then pointed out the possible outburst of small claims that first had not been observed.

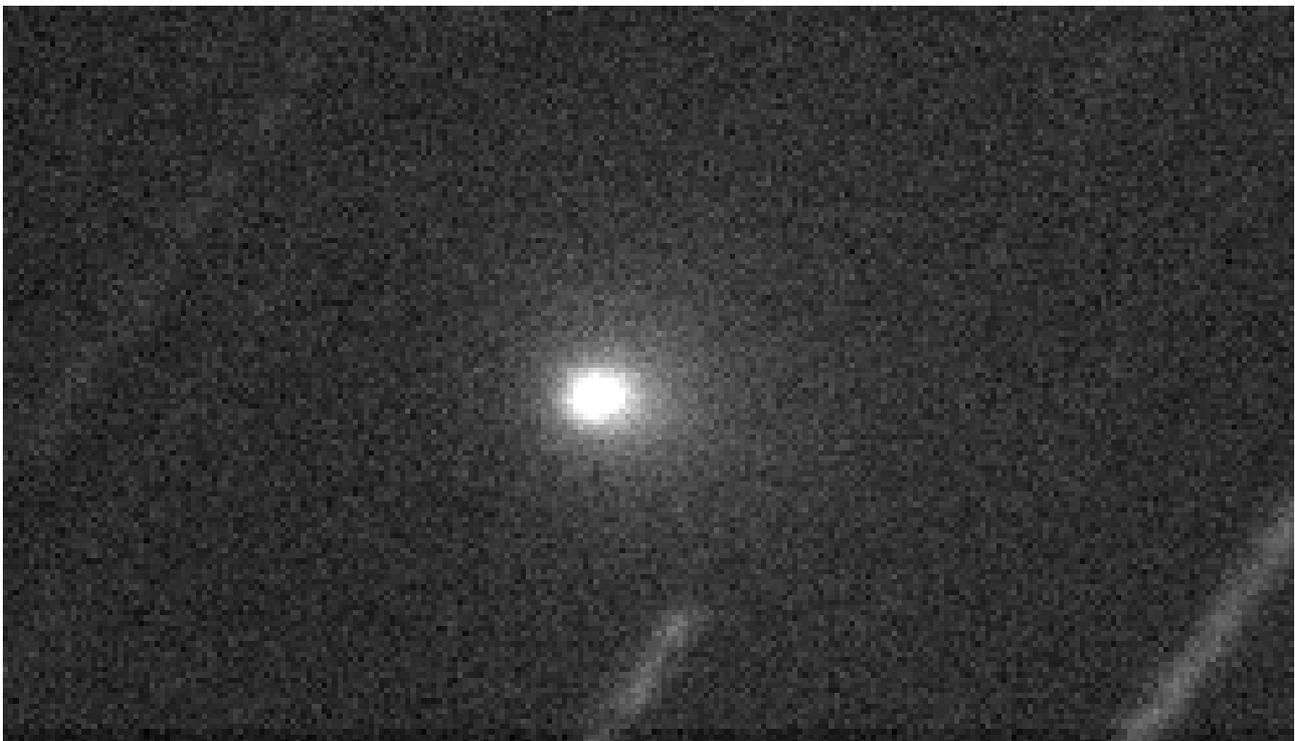

Fig. 10: Comet 6P observed on 25 August 2008 with a Newton Telescope 250 mm, 1200 mm focal length CCD camera and filter Rc

The increase in brightness was quite sudden, but apparently in agreement with his approach to our star, in August, reaching an apparent visual magnitude around 9. The survey CCD instead showed that there were signs of outburst that were not observed visually.

The series of images for our research were obtained on the evening of 25 August 2008, when the comet was 18 degrees over the location for our, DECL. = -26 45 10 south in the constellation of Capricorn.

There were obtained 120 images from 30 seconds each with a telescope Newton 250 mm aperture, focal length 1200 mm, Atik CCD filter and 16IC Rc direct focus.

(for the complete system see Tables 1 and 2).

The pursuit of hype was not optimal for different images are moved, but were the same used for granting the photometry. The inner part of the hair including false nucleus has been defocused to a FWHM of 14 pixels. (See fig. 11)

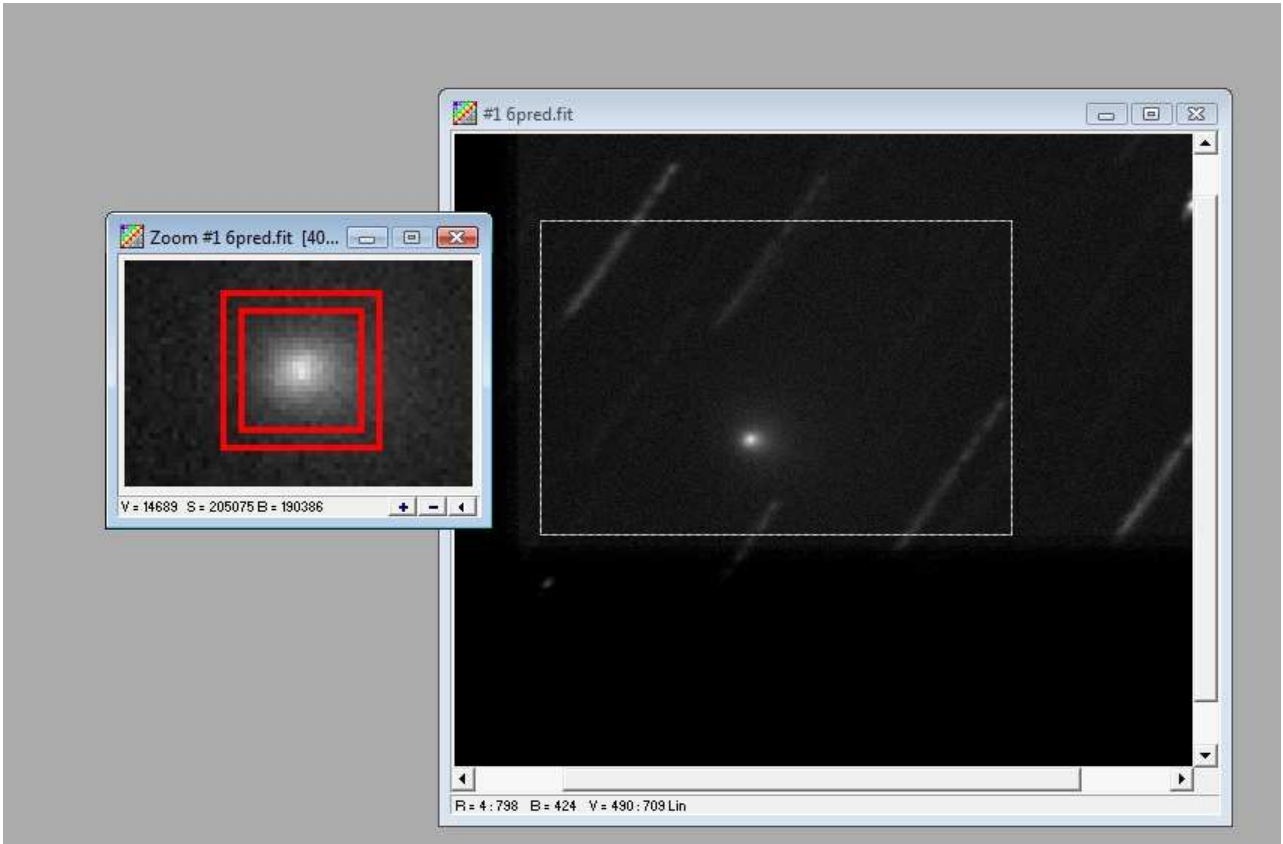

Fig. 11: Window aperture of 14 pixels for the inner side of the coma of the comet.

Even the stars in the field are therefore defocused. Those found to have good photometric are 5 and the listed in Table 8.

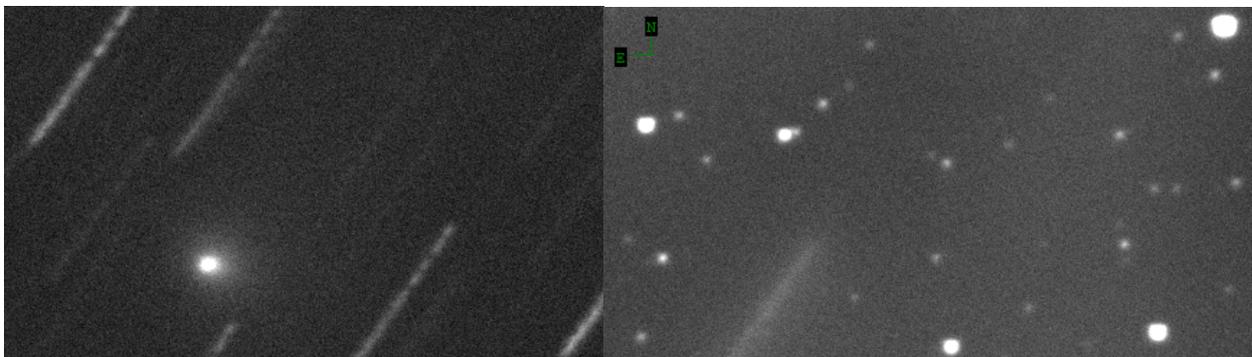

Fig. 12: Comet 6P observed the evening of 25 August 2008. On the left image was obtained with mediating Astroart 120 frames for 30 sec each, and aligning on the photometric centre of the comet. On the right the same images have been mediated by aligning on the centre of photometric stars. The error of alignment is 1-2 pixels.

| *Star/Catalogue* | mag. - USNO B1 | mag. - TYCHO 2 | mag. – GSC 2.2 |
|---|---|---|---|
| T 6916:1257:1(Ref1) | R1=10.53– R2=10.50 | Bt=11.223 – Vt=10.788 | |
| GSC 6916:1515 (Ref2) | R1=12.77 – R2=12.28 | | R=11.89 |
| GSC 6916:1396 (Ref3) | R1=12.88 - R2=12.61 | | R=12.19 |
| GSC 6916:1522 (Ref4) | R1=13.33 - R2=13.04 | | R=12.69 |
| GSC 6916:1391 (Ref5) | R1=13.46 – R2=12.93 | | R=14.41 |

Tab. 8: photometric parameters of the stars of reference derived from catalogues on-line using the Aladin, after calibrated astrometry the image aligned on the stars of the previous figure.

Photometric results obtained can be seen in the figures below.

Aperture 18 pixels - Resolution 1.26 arcsec / pixels - 25 cm telescope Newton f / 4.8 + CCD Atik 16IC + Rc

|  | Km/UA | In Meters |
|---|---|---|
| **Rho(Coma radius)** | 3556 | 3556000 |
| **D(Earth-Comet)** | 0,384000 | 57600000000 |
| **R(Sun-Comet)** | 1,359300 | 2,03895E+11 |

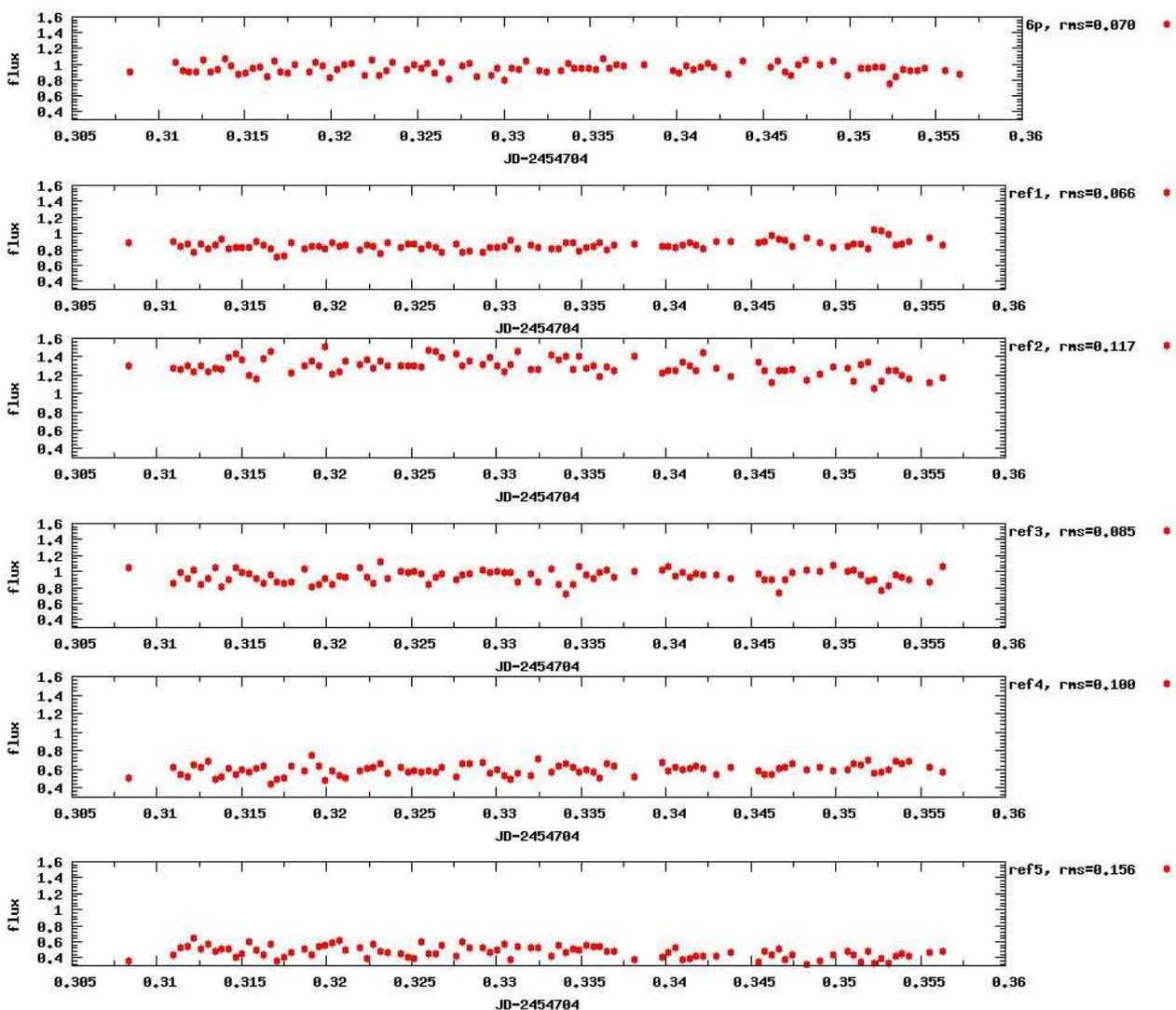

Fig. 13: The graphs show the flow of comet 6P compared with the flow of the stars of reference(see Table 8 for the characteristics of the stars). The **rms** respectively 0.070 for the comet and 0.066, 0.117, 0.085, 0.100 and 0.156 for the five stars of reference.

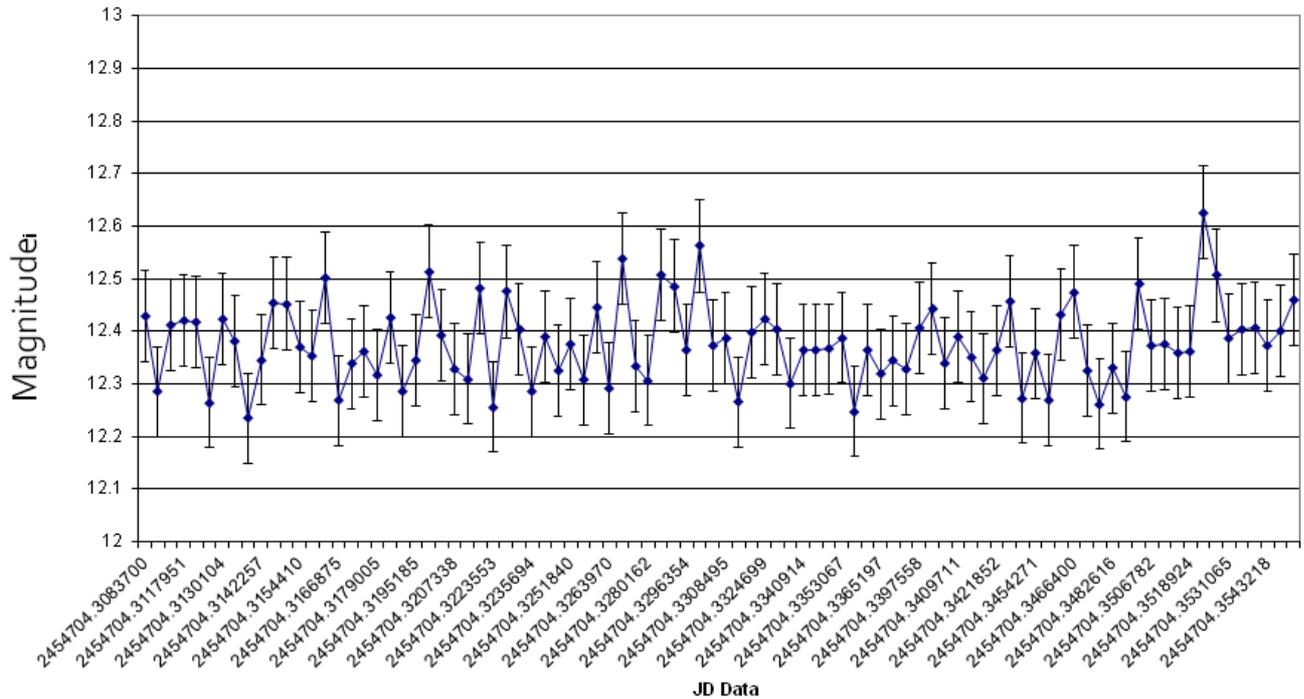

Fig. 14: The graph of the magnitude of the comet 6P with the flux converted into magnitudes.

Table 9 contains all of the counts of comet and star of reference for the complete set of images.

Tab. 9: flux of comet 6P and reference stars measured in ADU.

| JD | FLUX6P | REF1 | REF2 | REF3 | REF4 | REF5 |
|---|---|---|---|---|---|---|
| 2454704.3083700 | 37867 | 224178 | 47183 | 38927 | 20148 | 14576 |
| 2454704.3109861 | 42824 | 223268 | 45818 | 31885 | 23986 | 17024 |
| 2454704.3113912 | 39180 | 224716 | 46784 | 37926 | 21857 | 20870 |
| 2454704.3117951 | 39002 | 227117 | 48248 | 35008 | 20863 | 21727 |
| 2454704.3122002 | 40843 | 226255 | 48009 | 40420 | 27079 | 26528 |
| 2454704.3126042 | 45915 | 231501 | 49080 | 33373 | 25076 | 20758 |
| 2454704.3130104 | 40377 | 230241 | 47846 | 36436 | 28258 | 23665 |
| 2454704.3134155 | 41071 | 229023 | 48297 | 40692 | 20405 | 20019 |
| 2454704.3138194 | 46246 | 233043 | 46849 | 31612 | 20765 | 20386 |
| 2454704.3142257 | 43130 | 229366 | 52568 | 35948 | 25029 | 21327 |
| 2454704.3146308 | 38978 | 229414 | 53867 | 41207 | 22496 | 16884 |
| 2454704.3150359 | 39156 | 230281 | 51956 | 39110 | 24439 | 19139 |
| 2454704.3154410 | 42554 | 232979 | 46456 | 39065 | 23859 | 24958 |
| 2454704.3158472 | 41442 | 229932 | 43404 | 35045 | 24199 | 20071 |
| 2454704.3162836 | 37332 | 233405 | 52041 | 34386 | 26034 | 17982 |
| 2454704.3166875 | 46279 | 228677 | 54908 | 37957 | 18589 | 23658 |
| 2454704.3170914 | 42250 | 228710 | 52010 | 36268 | 21542 | 15776 |
| 2454704.3174965 | 41567 | 231089 | 49512 | 35910 | 22145 | 17856 |
| 2454704.3179005 | 39710 | 212056 | 42193 | 31064 | 23542 | 17604 |
| 2454704.3187095 | 39995 | 228884 | 49485 | 40509 | 23891 | 20922 |
| 2454704.3191134 | 45640 | 231753 | 51540 | 32687 | 30443 | 18268 |
| 2454704.3195185 | 41316 | 222134 | 47725 | 32381 | 24972 | 21428 |
| 2454704.3199236 | 36875 | 228439 | 56454 | 36080 | 19940 | 22840 |
| 2454704.3203287 | 40365 | 230221 | 45685 | 32720 | 23733 | 23736 |
| 2454704.3207338 | 43772 | 231807 | 47436 | 37165 | 22160 | 25331 |
| 2454704.3211389 | 43600 | 228490 | 50170 | 36296 | 20841 | 20299 |
| 2454704.3219491 | 38244 | 228196 | 50395 | 41505 | 24331 | 22020 |
| 2454704.3223553 | 46129 | 229800 | 51281 | 36517 | 24871 | 16364 |
| 2454704.3227604 | 40185 | 243119 | 51397 | 36123 | 27038 | 24729 |
| 2454704.3231644 | 43824 | 237051 | 55095 | 46937 | 29188 | 21709 |
| 2454704.3235694 | 45036 | 233168 | 48929 | 35965 | 22597 | 19059 |
| 2454704.3243773 | 41870 | 232529 | 50224 | 40205 | 26059 | 19267 |
| 2454704.3247789 | 44571 | 237753 | 50540 | 39775 | 24050 | 17214 |
| 2454704.3251840 | 42006 | 234547 | 49770 | 39477 | 24108 | 16752 |
| 2454704.3255880 | 48215 | 246234 | 53023 | 41294 | 25611 | 26804 |
| 2454704.3259931 | 40094 | 237541 | 56203 | 34457 | 24437 | 19020 |
| 2454704.3263970 | 45747 | 232447 | 55212 | 36998 | 23827 | 18972 |
| 2454704.3268021 | 37021 | 229948 | 53939 | 39498 | 26476 | 23454 |
| 2454704.3276111 | 44084 | 237780 | 54585 | 36361 | 22018 | 17655 |
| 2454704.3280162 | 47151 | 236811 | 52473 | 40086 | 28361 | 26275 |
| 2454704.3284201 | 39204 | 237875 | 54199 | 40514 | 28482 | 23081 |
| 2454704.3292303 | 39980 | 236834 | 52838 | 42414 | 29297 | 22900 |
| 2454704.3296354 | 43274 | 234596 | 53848 | 39957 | 23576 | 20152 |
| 2454704.3300405 | 37443 | 243613 | 52748 | 41703 | 26235 | 21922 |
| 2454704.3304456 | 44254 | 244385 | 50139 | 40886 | 23394 | 24578 |
| 2454704.3308495 | 42377 | 243936 | 51130 | 39786 | 20902 | 16257 |
| 2454704.3312546 | 46943 | 231648 | 55593 | 35349 | 23199 | 22439 |
| 2454704.3320648 | 42695 | 242374 | 50438 | 40231 | 22914 | 22850 |
| 2454704.3324699 | 42816 | 245335 | 51396 | 37125 | 31012 | 23130 |
| 2454704.3332801 | 42389 | 237410 | 55330 | 42015 | 24349 | 18268 |
| 2454704.3336863 | 46439 | 237130 | 53607 | 34724 | 26854 | 24042 |
| 2454704.3340914 | 44372 | 246804 | 55613 | 30492 | 28332 | 20285 |
| 2454704.3344977 | 43895 | 244160 | 50004 | 34655 | 26424 | 22092 |
| 2454704.3349016 | 44738 | 237793 | 56328 | 43824 | 25043 | 21970 |
| 2454704.3353067 | 43563 | 241437 | 51260 | 39982 | 25582 | 24261 |
| 2454704.3357106 | 49113 | 241942 | 51663 | 37861 | 24389 | 23251 |
| 2454704.3361157 | 43782 | 243737 | 47455 | 40399 | 21849 | 22947 |
| 2454704.3365197 | 46067 | 237199 | 51358 | 41909 | 28308 | 20642 |
| 2454704.3369236 | 44631 | 240665 | 49711 | 38341 | 27084 | 20433 |
| 2454704.3381366 | 45260 | 241711 | 55006 | 40648 | 22410 | 16245 |
| 2454704.3397558 | 43172 | 245454 | 49875 | 42573 | 29218 | 18015 |
| 2454704.3401609 | 40415 | 237169 | 49326 | 42724 | 24539 | 19745 |
| 2454704.3405660 | 44761 | 238140 | 49681 | 38713 | 26481 | 22555 |
| 2454704.3409711 | 41853 | 236062 | 51351 | 39448 | 24890 | 16242 |
| 2454704.3413762 | 44323 | 243477 | 51445 | 38304 | 25997 | 17168 |
| 2454704.3417801 | 45337 | 238029 | 49023 | 39085 | 26613 | 17749 |
| 2454704.3421852 | 43278 | 233822 | 55607 | 38472 | 25389 | 18162 |
| 2454704.3429954 | 39533 | 241041 | 49570 | 38782 | 22750 | 17855 |
| 2454704.3438067 | 47807 | 245839 | 47268 | 37791 | 26391 | 20100 |
| 2454704.3454271 | 43857 | 242345 | 52525 | 39372 | 24995 | 15223 |
| 2454704.3458310 | 49617 | 254965 | 51602 | 38497 | 24204 | 21322 |
| 2454704.3462361 | 46831 | 287115 | 51169 | 42109 | 26491 | 21288 |
| 2454704.3466400 | 40746 | 256219 | 51052 | 31434 | 26603 | 22317 |
| 2454704.3470451 | 46335 | 252329 | 50579 | 37702 | 26808 | 16425 |
| 2454704.3474537 | 47700 | 237153 | 49662 | 39809 | 27608 | 18685 |
| 2454704.3482616 | 43618 | 240948 | 44360 | 39867 | 24539 | 13244 |
| 2454704.3490683 | 46483 | 239310 | 47070 | 40047 | 25990 | 15664 |
| 2454704.3498715 | 39114 | 239155 | 50893 | 43584 | 24719 | 18817 |
| 2454704.3506782 | 43561 | 239541 | 50671 | 41086 | 25429 | 20481 |
| 2454704.3510833 | 43571 | 243364 | 45814 | 41517 | 28270 | 19256 |
| 2454704.3514873 | 44280 | 243450 | 52107 | 39636 | 27828 | 15329 |
| 2454704.3518924 | 44370 | 239198 | 53437 | 36852 | 29624 | 21106 |
| 2454704.3522975 | 41430 | 311198 | 51473 | 44277 | 28552 | 17169 |
| 2454704.3527025 | 42537 | 284545 | 50191 | 35616 | 27059 | 19081 |
| 2454704.3531065 | 45634 | 268876 | 52894 | 36145 | 26812 | 15087 |
| 2454704.3535116 | 41079 | 233942 | 48088 | 38004 | 28196 | 17865 |
| 2454704.3539167 | 41866 | 241125 | 47377 | 38085 | 27682 | 19558 |
| 2454704.3543218 | 43037 | 242980 | 45975 | 36784 | 28496 | 18200 |
| 2454704.3555359 | 41752 | 245819 | 44444 | 35270 | 26091 | 19685 |
| 2454704.3563449 | 40016 | 240343 | 46839 | 42883 | 24396 | 20637 |

The median values derived with the fluxes of table 9 are the follows:

**magstar = 10,029 (+ / -0.024), magcomet = 12,370 (+ / -0.077).**

Looking to the Light Curves in fig.5, fig.9 and fig.14 it is clear that defocusing of comets CCD images to improve the photometry is not the better way if the S/N ratio is not high. The precision don't increase and the error is about 2 time the error calculated on photometry without defocus.

By the way, with photometry differential on focused images of comets it is possible to detect variations of magnitude at least on the order of 2%-3% of magnitude. This implied that on a time series of images of 2-4 hours of observation, it is possible to measure variations of brightness possibly linked with the rotation of the nucleus and/or with short and minor outburst.

## Acknowledgements.
M. Barbieri for usefull hints and extrasolar planets photometry's theory explanation.

___________________________________